\documentclass[aps,prl, preprintnumbers,groupedaddress,nofootinbib,twocolumn]{revtex4-1}

\pdfoutput=1
\usepackage{graphicx}
\usepackage{latexsym}
\usepackage{amsfonts}
\usepackage{amssymb}
\usepackage{amsmath}
\usepackage{slashed}
\usepackage{feynmp}
\usepackage{hyperref}
\usepackage{url}
\usepackage{color}
\usepackage{cancel}
\usepackage{multirow,array}

\begin{document}

\title{Boosting the Direct CP Measurement of the Higgs-Top Coupling}

\author{Matthew R.~Buckley}
\affiliation{Department of Physics and Astronomy, Rutgers University, Piscataway, NJ 08854, USA}
\author{Dorival Gon\c{c}alves}
\affiliation{Institute for Particle Physics Phenomenology, Department of Physics, Durham University, UK}

\begin{abstract}
Characterizing the 125~GeV Higgs  is a critical component of the physics program at the LHC \mbox{Run II}. 
In this Letter, we consider $t\bar{t}H$ associated production in the dileptonic mode.
We demonstrate that the difference in azimuthal angle between the leptons from top decays can directly reveal 
the CP-structure of the top-Higgs coupling with the sensitivity of the measurement substantially enhanced in the boosted Higgs 
regime.  We first show how to access this channel via $H \to b\bar{b}$ jet-substructure tagging, then demonstrate the ability 
of the new variable to measure CP. Our analysis includes all signal  and background samples simulated via the \textsc{MC@NLO} 
algorithm including hadronization and underlying-event effects. Using boosted Higgs substructure with dileptonic tops, we find that 
the top-Higgs coupling strength  and the CP structure can be directly probed with achievable luminosity at the 13~TeV LHC.
\end{abstract}

\maketitle

Determining the properties of the Higgs particle $H$ at 125~GeV will provide important information about the
as-yet unknown physics beyond the Standard Model (SM), and is therefore an important focus of the LHC Run II. 
Presently its couplings to $W$ and $Z$ gauge bosons are directly measured through the Higgs decays to vector 
boson pair  and are consistent with a spin-0 particle with SM-strength  CP-even 
couplings~\cite{higgs,discovery,lhc_runI,legacy,Plehn:2001nj}. However, the ratios between scalar and pseudoscalar 
couplings  might differ from channel to channel in the presence of CP violation. Hence, it is of fundamental importance to 
access this information in as many channels as possible.\footnote{CP-odd Higgs-vector boson couplings
can appear only through operators of dimension-6 or higher~\cite{dim6},  while CP-odd Higgs-fermion 
couplings could manifest at tree level. Thus, the latter  are naturally more sensitive to CP  violation than the
former.} Of particular interest is the coupling to top quarks, as $y_t^{SM}\sim{\cal O}(1)$.

The strength and CP-structure of the top-Higgs coupling  are currently inferred from the measured Higgs-gluon and Higgs-photon
interactions  through the production $gg \to H$ and decay $H \to \gamma\gamma$ channels~\cite{Djouadi:2013qya,Klamke:2007cu}, 
as well as constraints on electron dipole moments \cite{Brod:2013cka}.  However, 
as these couplings are loop-induced, the measurements could be a combination of SM and new physics~\cite{loop_ind,Arbey:2014msa}. 
Direct measurements of both the strength and CP-properties of this coupling are necessary to disentangle new 
physics  effects. The associated Higgs with $t\bar{t}$ pair production qualifies as the most direct probe.

In this Letter, we demonstrate that the $t\bar{t}H$ channel can be measured with dileptonic top pairs and 
Higgs decay to $b\bar{b}$ via jet substructure~\cite{bdrs,tth_plehn,Seymour:1993mx} (to our knowledge, this Letter is the first to use
 boosted Higgs  substructure associated to dileptonic top pair). Including higher order QCD effects to 
signal and backgrounds via the \textsc{MC@NLO} algorithm~\cite{mcatnlo}, we show that this channel can be probed with
a reasonable luminosity in the Run II  LHC.  In the same channel we then consider the direct CP measurement of the Higgs-top coupling via spin 
correlations. The lab frame CP-sensitive variable we propose is $\Delta\phi_{\ell\ell}$:  the difference in azimuthal angle around the beam axis of the top pair decay leptons. 
This is somewhat similar to observables proposed in previous  works~\cite{Ellis:2013yxa,Boudjema:2015nda,tth_NLO,Biswas:2014hwa, 
Casolino:2015cza,Kolodziej:2015qsa}. However, the CP-sensitivity of $\Delta\phi_{\ell\ell}$ is enhanced at large Higgs 
transverse  momentum $p_{TH}$. Fortunately for our purposes, this requirement 
dovetails nicely with the kinematic  region required for jet substructure Higgs tagging.  Thus, high-$p_{TH}$ dileptonic $t\bar{t}H$ events
have experimentally  attractive properties both for initial discovery and CP-structure measurement.\medskip

%----------------------------------------------------------------------------------------
%\section{Spin correlations in $t\bar{t}H$ decays}

We parametrize the  top-Higgs interaction as
%---------------------
\begin{equation}
{\cal L} \supseteq 
-\frac{m_t}{v} K \bar{t}\left( \cos\alpha +i \gamma_5 \sin\alpha \right) t~H,
\label{eq:topcoupling}
\end{equation}
%---------------------
where $K$ is a real number and  $\alpha$ a CP-phase. The CP-even SM Higgs $0^+$ particle is $(K,\alpha) = (1,0)$, while $\alpha = \tfrac{\pi}{2}$ corresponds to a CP-odd $0^-$. 

%----------------------------------------------------------------------------------------
\begin{figure*}[ht!]
\includegraphics[width=0.5\columnwidth]{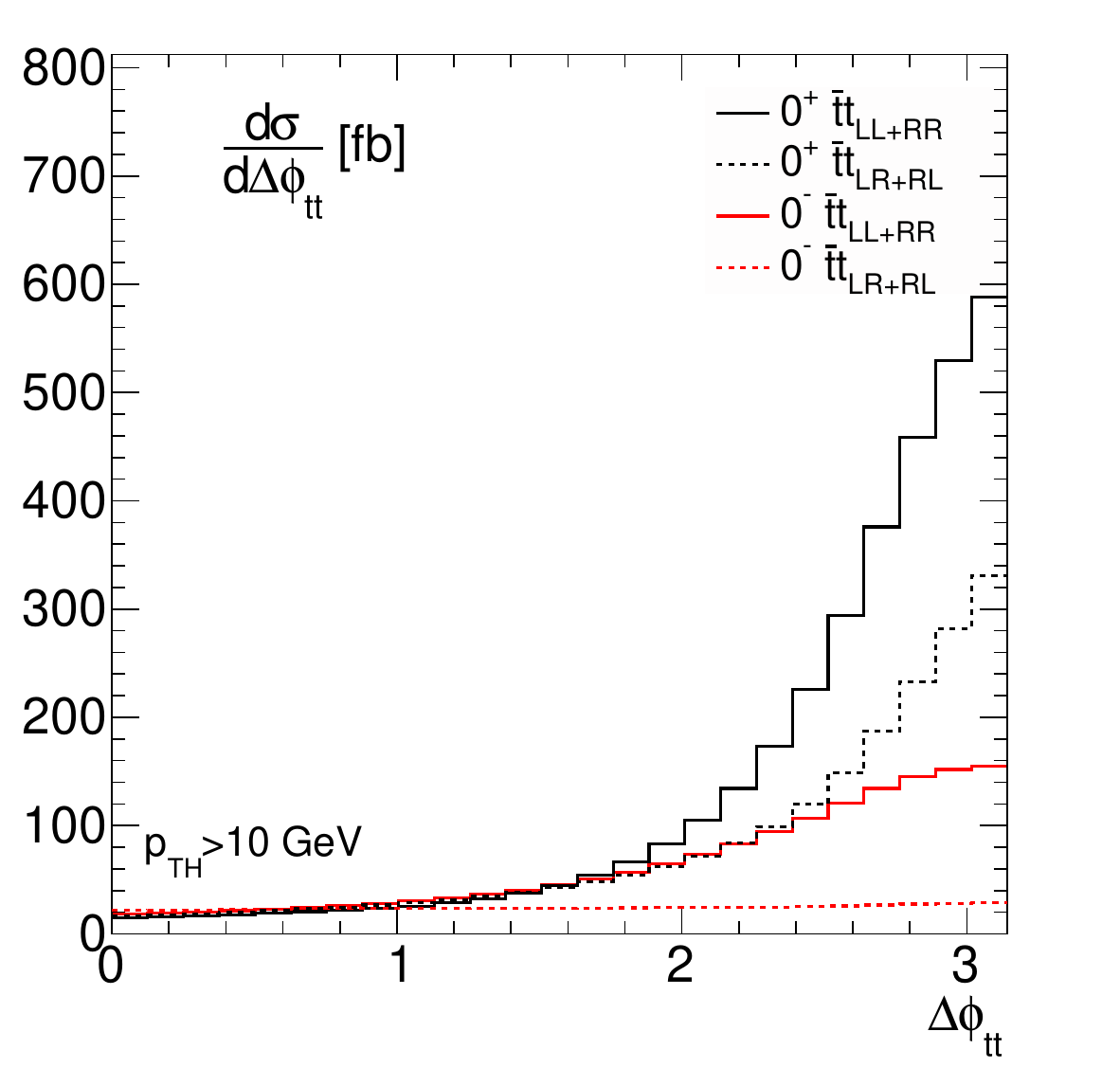}
\includegraphics[width=0.5\columnwidth]{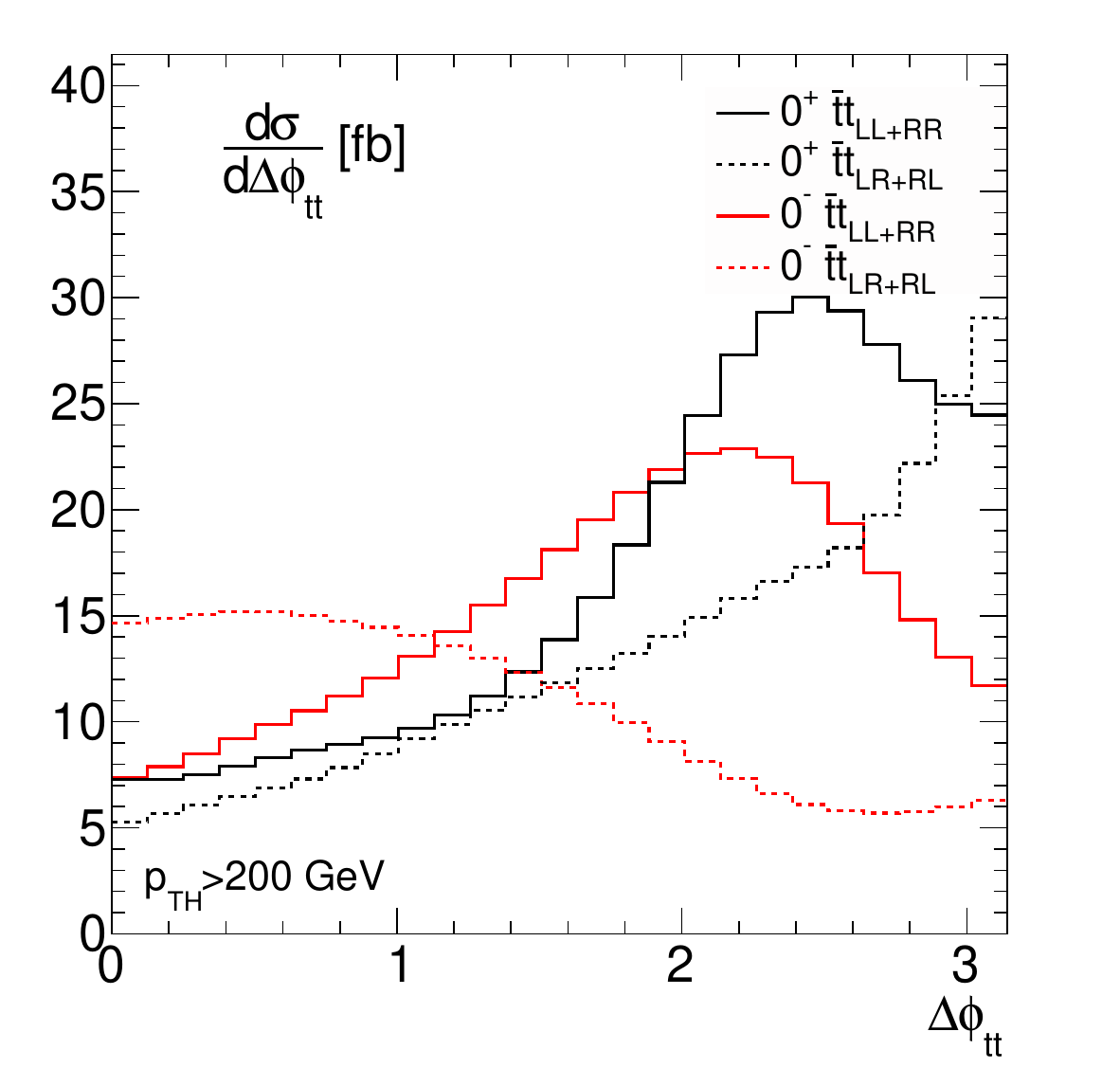}
\includegraphics[width=0.5\columnwidth]{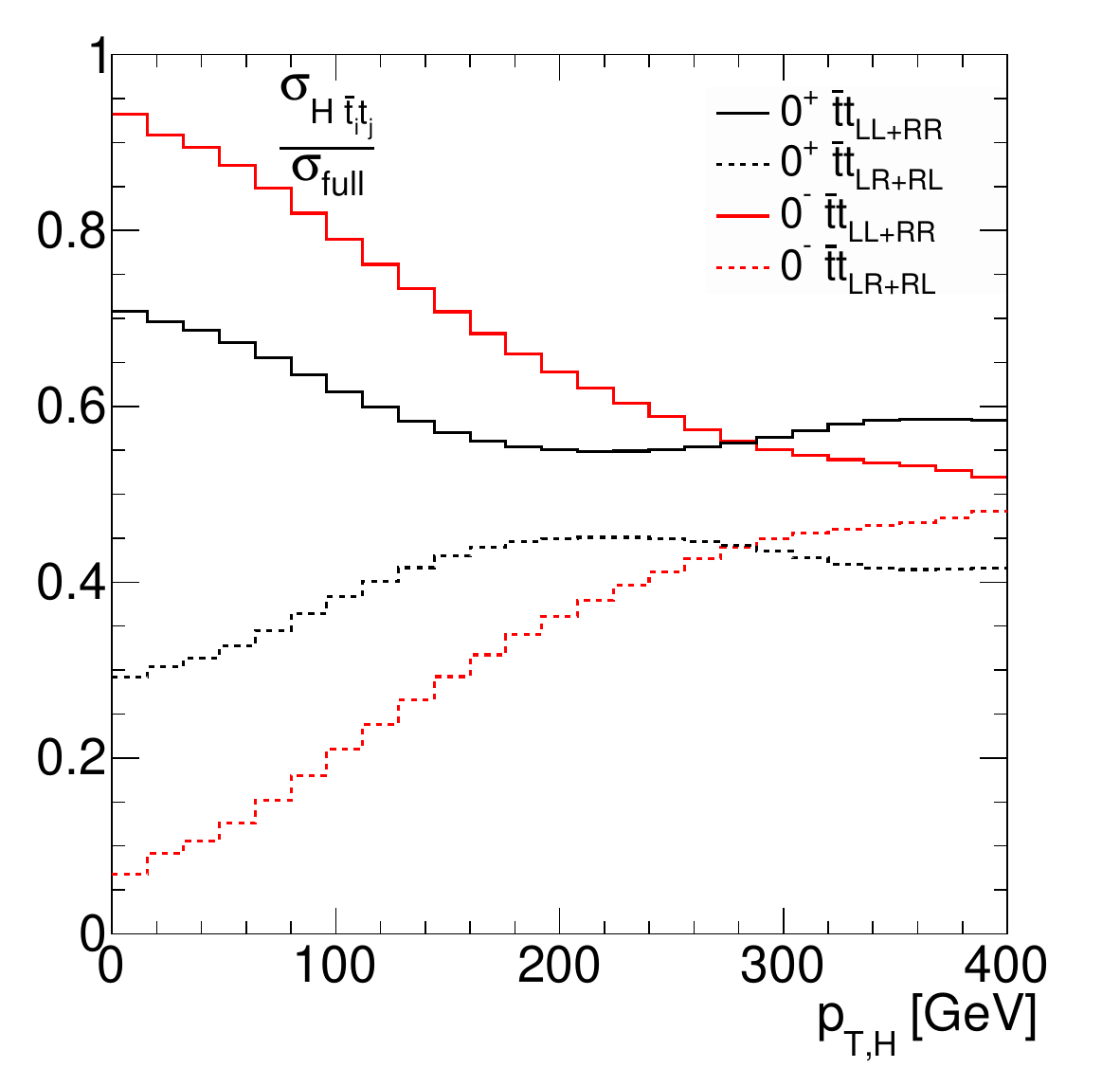}
\includegraphics[width=0.5\columnwidth]{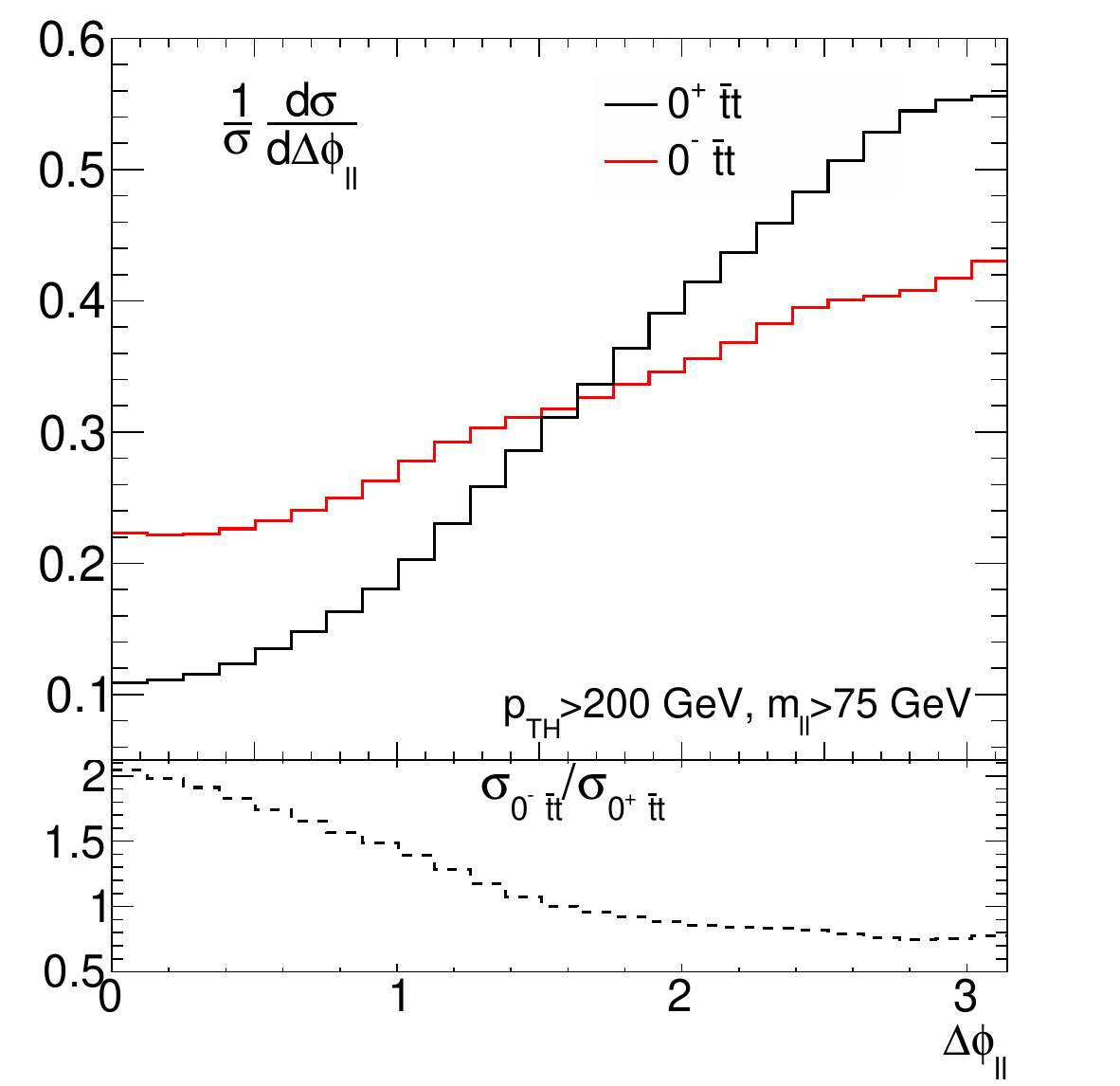}
\caption{Left and Left-Center:  $\Delta\phi_{tt}$  distribution for CP-even $0^+$ (black) and
CP-odd $0^-$ (red) couplings  without requiring a boosted Higgs (left panel) and in
the boosted regime $p_{TH}> 200$~GeV (left-center).  Contributions from the like-helicity
 $t_L\bar{t}_L+t_R\bar{t}_R$ and unlike-helicity $t_L\bar{t}_R+t_R\bar{t}_L$ states are shown in
 solid and dashed lines, respectively.  Right-Center: Fraction of  like-helicity  and unlike-helicity  states 
 as a function of the minimum Higgs transverse momentum selection cut $p_{T,H}$. Right: $\Delta\phi_{\ell\ell}$ 
 parton level distribution for CP-even and CP-odd Higgs  with $p_{TH} > 200$~GeV and $m_{\ell\ell}>75$~GeV.
 \label{fig:deltaphitt}}
\end{figure*}
%----------------------------------------------------------------------------------------

In principle, the anatomy of the top-Higgs interaction can be revealed via spin correlations, both at the LHC~\cite{Ellis:2013yxa,Boudjema:2015nda} and a future $e^+e^-$ collider \cite{BhupalDev:2007is}. In the other LHC-focused works, the proposed variable's sensitivity is washed out by experimentally required selection criteria.
Analogously to the $t\bar{t}$ production  studied in Ref.~\cite{Mahlon:1995zn}, distinct kinematic distributions exist in $t\bar{t}H$ 
production between the like-helicity  $(t_L\bar{t}_L+t_R\bar{t}_R)$ and unlike-helicity 
$(t_L\bar{t}_R+t_R\bar{t}_L)$ top pairs. We adopt helicity conventions as in \textsc{HELAS}~\cite{helas}. 
 
For our analytic argument, we will consider the distribution of top-pairs in the $t\bar{t}H$ production. Without full top-quark reconstruction, such distributions are not directly accessible. However, the spin-correlations between the top pairs are passed on to the top decay products, %where the decay products $t\rightarrow W^{+} +b$ 
%with $W^{+}\rightarrow l^+\nu~(\bar{d}u)$  
which are correlated with the top spin axis. The charged lepton and $d$-quark from the $W$-boson decay have the largest degree of correlation  with the top quark spin axis~\cite{Mahlon:1995zn}. Hence experimentally accessible leptonic angular variables from dileptonic top decay (such as $\Delta\phi_{\ell\ell}$ which we will demonstrate contains CP-information), can be used as an experimentally clean proxy for the more fundamental variables built from $t$ and $\bar{t}$ momenta, such as $\Delta\phi_{tt}$, considered here.

%Having seen that we should be interested in the highly-boosted, mixed-helicity $t\bar{t}H$ events, we now turn to the analytic argument for this behavior. 

To further simplify our analysis, we focus on the $q\bar{q}$-initiated $s$-channel production of the  $t\bar{t}H$ state,
%. Due to the falling gluon parton 
%distribution function, such initial  states are more important at large $p_{TH}$, which is the kinematic regime the experiments will be forced into by selection cuts 
though a nearly identical argument follows when considering $s$-channel gluon-gluon production. 
We further restrict our consideration to top-antitop pairs of mixed helicity, $t_L\bar{t}_R H$ and $t_R\bar{t}_L H$, which transform into 
themselves  under CP. These apparently arbitrary choices will be justified shortly. %We know from Fig.~\ref{fig:deltaphitt} that these states will have the greatest sensitivity to the angle $\alpha$.

With incoming quark and antiquark momenta $q_1$ and $q_2$, outgoing top and antitop 
momenta $k_1$, $k_2$, and Higgs momentum $p$, the mixed helicity state matrix element is
%---------------------
\begin{eqnarray}
{\cal M} & \propto & \frac{m_t\left[\bar{v}(q_2)\gamma^\mu u(q_1) \right] 
\left[ \bar{u}(k_1)P_{L/R}A\gamma_\mu P_{R/L} v(k_2)\right]}
{\left[ q_1+q_2\right]^2\left[m_H^2+2k_1\cdot p \right]\left[ m_H^2+2k_2\cdot p \right]}  
\;,
\label{eq:Mdef}
 \\
% A &= & (m_H^2+2k_2\cdot p)e^{i\alpha \gamma_5}+(m_H^2+2k_1\cdot p)e^{-i\alpha\gamma_5}.
A& =& \left[\tfrac{m_H^2}{2}+(k_1+k_2)\cdot p\right]\cos\alpha-i\left[(k_1-k_2)\cdot p \right]\gamma_5\sin\alpha. \nonumber
\end{eqnarray}
%---------------------
The matrix $A$ is the only source of a possible kinematic difference resulting from the CP-structure of the top-Higgs coupling. 
Of course, $\Delta\phi_{tt}$ (as with many other kinematic variables) will appear in other locations in the matrix element and 
phase space factors, but any kinematic difference arising from $\alpha$ must come from $A$. 

%We can rewrite $A$ as
%%---------------------
%\begin{equation}
%A = \left[\tfrac{m_H^2}{2}+(k_1+k_2)\cdot p\right]\cos\alpha-i\left[(k_1-k_2)\cdot p \right]\gamma_5\sin\alpha.
%\label{eq:Adef}
%\end{equation}
%%---------------------
The ideal set of kinematic variables measuring $\alpha$ in the mixed helicity top final state is one that is maximally sensitive to both 
$(k_1+k_2)\cdot p$ and $(k_1-k_2)\cdot p$. Unfortunately, using these kinematic combinations directly requires full event reconstruction, which is challenged by jet energy uncertainties and missing energy in the leptonic decays~\cite{Buckley:2013auc}. 
 Our chosen variable, $\Delta\phi_{tt}$, inhabits a happy medium, probing $(k_1\pm k_2)\cdot p$ as we will show, while being closely related 
 the easily measured $\Delta\phi_{\ell\ell}$. 

%----------------------------------------------------------------------------------------
%\begin{figure}[t!]
%\centering
%\includegraphics[width=0.8\columnwidth]{./dphill_pth200.pdf}
%\caption{ $\Delta\phi_{\ell\ell}$ parton level distribution for CP-even and CP-odd Higgs couplings with $p_{TH} > 200$~GeV 
%and $m_{\ell\ell}>75$~GeV.}
% \label{fig:deltaphill}
%\end{figure}
%----------------------------------------------------------------------------------------

The dependence on $\Delta\phi_{tt}$ in the coefficients of Eq.~\eqref{eq:Mdef} is maximized in the high-momentum regime. Performing a boost 
along the beam axis (which leaves $\Delta\phi_{tt}$ unchanged) to the frame where the Higgs is perpendicular to the beam, the coefficients can
 be written in terms of the sum of the top-antitop azimuthal angles $\Sigma_{tt}$ and their difference $\Delta\phi_{tt}$. We see that we can approximate
%---------------------
\begin{eqnarray}
(k_1+k_2)\cdot p & \propto & \sin\left(\tfrac{\Delta\phi_{tt}}{2}\right)\cos\left(\tfrac{\Sigma_{tt}}{2}\right)-\tfrac{1}{2}\sin\Delta\phi_{tt}, \label{eq:kplus} \\
(k_1-k_2)\cdot p & \propto & \cos\left(\tfrac{\Delta\phi_{tt}}{2}\right)\sin\left(\tfrac{\Sigma_{tt}}{2}\right)-\tfrac{1}{2}\sin\Sigma_{tt}. \label{eq:kminus}
\end{eqnarray}
%---------------------
The key observation here is that the CP-even couplings oscillates with sines of $\Delta\phi_{tt}$, and the CP-odd couplings with cosines. 
Integrating over $\Sigma_{tt}$, we see that the CP-even (odd) coupling has a deficit of events at $\Delta\phi_{tt} = 0(\pi)$ and 
an excess at $\pi(0)$. The form of this result can
 also be obtained by considering the interference between spin-states~\cite{Buckley:2007th}, and requiring that the mixed helicity
  states transform as $(-1)^j$ for the CP-even couplings and $(-1)^{j+1}$ for the CP-odd, for total angular momentum $j$.

This analytic argument is borne out in simulation, using the full matrix element calculation, all initial state partons (not just quark/antiquarks), and summing over all helicities. Fig.~\ref{fig:deltaphitt} shows the differential distribution $\Delta\phi_{tt}$ with and without the large Higgs $p_{TH}$ cut. 
At low $p_{TH}$, $\Delta\phi_{tt}$ has a minimum at  $\sim 0$ and peaks at $\sim \pi$ for both CP-even and CP-odd couplings.
However, at high $p_{TH}$ regime and in the unlike-helicity $t\bar{t}H$ final states, the CP-sensitivity of $\Delta\phi_{tt}$ becomes clear. As in our analytic argument, these helicity 
 combinations develop peaks at $\Delta\phi_{tt}\sim 0$ and a minimum at $\sim \pi$ for the CP-odd coupling, opposite to the distributions for the
other final states.  Fortunately, the high $p_{TH}$ regime also enhances the signal-containing mixed helicity configuration~\cite{Biswas:2014hwa}. 
As seen in Fig.~\ref{fig:deltaphitt} (right-center), the unlike-helicity fraction goes from $\sim7\%$ $(30\%)$ of the cross-section at low $p_{TH}$ selection
to  $\sim40\%$ ($45\%$) at  $p_{TH}>200$~GeV for the CP-odd (even) state.

The right panel of Fig.~\ref{fig:deltaphitt} shows the differential $\Delta\phi_{\ell\ell}$ distribution in the CP-even and CP-odd scenarios
with $p_{TH} > 200$~GeV and $m_{\ell\ell}>75$~GeV.  This second requirement is a proxy for $m_{tt}$, further enhancing the 
unlike-helicity final states~\cite{Mahlon:1995zn}. As with $\Delta\phi_{tt}$, the behavior of the $0^-$ coupling is clearly distinguished from the $0^+$
assumption by an increase in events near $\Delta\phi_{\ell\ell} \sim 0$ and a deficit near $\pi$.

%Having constructed the analytic argument, we return to the results of Fig.~\ref{fig:deltaphitt}. %where we show the differential cross section with 
%respect to $\Delta\phi_{tt}$ at the parton-level, both without and with a cut on the $p_T$ of the Higgs. 
%Without a cut on $p_{TH}$, 
%the signal-containing mixed helicity states are subdominant to same-helicity top final states. Requiring the Higgs to
%be boosted increases the percentage of mixed helicity tops~\cite{Biswas:2014hwa}, enhancing the dependence on the CP structure of the 
%top-Higgs coupling as measured by $\Delta\phi_{tt}$ and $\Delta\phi_{\ell\ell}$, see Fig.~\ref{fig:deltaphitt}.

Requiring  Higgs $p_{TH}>200$~GeV is a sacrifice of total cross section. However, in our analysis, 
we consider the $t\bar{t}H$ channel with dileptonic top decay and Higgs decay to $b\bar{b}$. As we will describe, jet-substructure
tagging can be used in this channel to distinguish signal from background. Requiring collimated $b$-quarks in a fat-jet~\cite{bdrs,tth_plehn}
implies a large boost for the Higgs. Our CP-sensitive signal is enhanced with the same kinematics required for background rejection.

We now
%Now that we have developed a lab-frame variable built from simple experimentally measurable quantities 
%which directly probes the CP-structure of the $t\bar{t}H$ coupling, we 
turn to the question of realistic event 
selection, background rejection, and required luminosity. We first show that we can assess this channel
at the Run II LHC and then that we can directly probe its CP structure. \medskip

%%----------------------------------------------------------------------------------------
%\section{Analysis}

We consider the Higgs-top in $pp\rightarrow t\bar{t}H$ with dileptonic tops
and the Higgs boson decay $H\rightarrow b\bar{b}$ at the ${\sqrt{s}=13}$~TeV LHC. We demand
four bottom tagged jets and two opposite-sign leptons. The main backgrounds for this
process in order of relevance are  $pp\rightarrow t\bar{t}b\bar{b}$ and $t\bar{t}Z$.
 
The signal $t\bar{t}H$ sample is  generated with  \textsc{MadGraph5}+\textsc{Pythia8}~\cite{mg5,pythia8}, and the 
$t\bar{t}b\bar{b}$  and  $t\bar{t}Z$ backgrounds with \textsc{Sherpa+OpenLoops}~\cite{sherpa,openloops}.  All signal and 
background samples  are simulated with the~\textsc{MC@NLO} algorithm~\cite{mcatnlo} and account for hadronization
and underlying event effects. Since the Higgs boson is part  of a multi-jet system, a proper modeling of QCD  effects is 
fundamental in this study. Hence, we include the higher order QCD contributions to all considered processes. 

Next-to-leading-order (NLO) process generation requires factorization between the $t\bar{t}H$ production and decays.
Spin correlations are restored in our simulations by \textsc{MadSpin}~\cite{madspin} and the respective~\textsc{Sherpa}
module~\cite{sherpa_spin}. Their output were in agreement at leading order with a full decay chain simulation.\medskip

Our search strategy relies on the background suppression at the boosted regime~\cite{bdrs,tth_plehn,Seymour:1993mx},
that opportunely enhances the desired spin correlation effects, as previously mentioned. We start our analysis 
with some basic leptonic selections: two isolated opposite-sign leptons  with ${p_{T\ell}>15}$~GeV 
and $|\eta_\ell|<2.5$. %The lepton is isolated if the transverse energy deposited within a  cone $R=0.2$ around the
%lepton is  less than  $20\%$ of its transverse energy. 
The hadronic part of the event uses the Cambridge/Aachen (C/A) jet algorithm~\cite{fastjet}, and requires at least one  
boosted $({p_{TJ}>200}~\text{GeV})$ and central $(|\eta_J|<2.5)$ fat-jet ${(R =1.2)}$. This must be Higgs-tagged via the 
BDRS algorithm~\cite{bdrs,tth_plehn}, requiring three subjets where the two hardest are $b$-tagged. 
We assume $70\%$ $b$-tagging efficiency and 1\% mistag rate~\cite{btagging}. Possible pileup effects on the Higgs
mass are controlled by the BDRS filtering, as it has been  shown on LHC data~\cite{pileup}.  After a successful Higgs tag, 
we remove the Higgs fat-jet from the event and re-cluster the remaining hadronic activity with C/A using a smaller jet radius 
$R=0.5$. As the signal does not have any additional high mass particle decaying hadronically, we can safely suppress 
the underlying event contamination by decreasing the jet size. We then demand at least  two  jets with $p_{Tj}>30$~GeV 
and $|\eta_j|<2.5$, at least two of which $b$-tagged.

%----------------------------------------------------------------------------------------
\begin{figure}[t!]
\centering
\includegraphics[width=0.9\columnwidth]{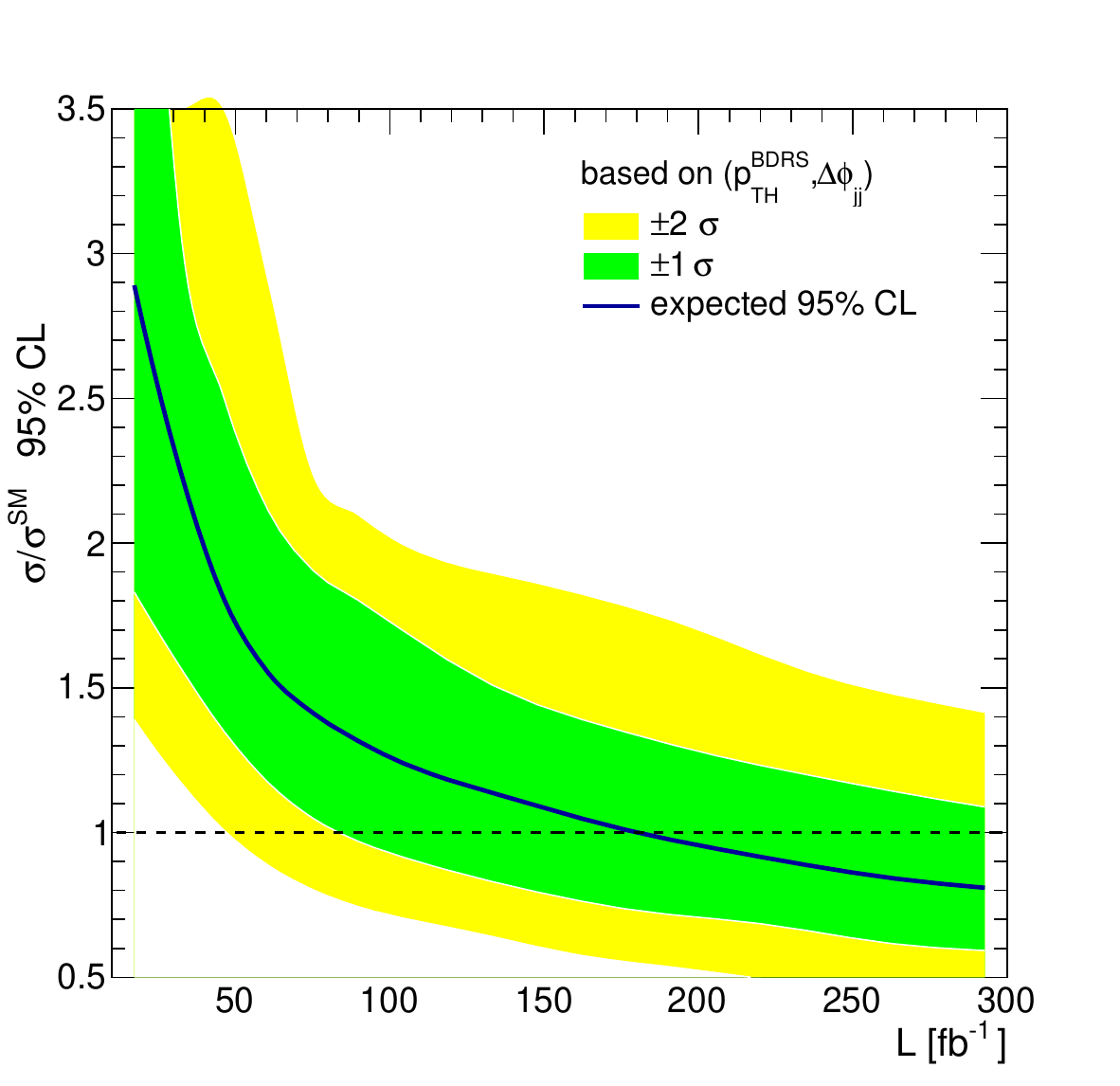}
\caption{Expected 95\% CL upper limits on $\sigma/\sigma^{SM}$ for four $b$-tag dileptonic $t\bar{t}(H\to b\bar{b})$
 as a function of LHC luminosity.} 
 \label{fig:luminosity}
\end{figure}
%----------------------------------------------------------------------------------------

To enhance signal/background ratio and suppress combinatorics, we require that the reconstructed
mass for the filtered Higgs to be in the window ${|m_H^{BDRS}-m_H|<10}$~GeV and the filtered $b$-tagged jets to
have $m_{b\bar{b}}>110$~GeV. The detailed cut-flow is presented in Table~\ref{tab:cut_flow}. 

%-------------------------------------------------------
\begin{table}[h!]
\centering
\begin{tabular}{l | c  | c | c  }
  \multicolumn{1}{c|}{cuts} &
  \multicolumn{1}{c|}{$t\bar{t}H$}&
  \multicolumn{1}{c|}{$t\bar{t}b\bar{b}$} &
  \multicolumn{1}{c}{$t\bar{t}Z$}
   \\
  \hline
{BDRS $H$-tag, $p_{T\ell}>15$~GeV,  $|\eta_\ell|<2.5$}   
&\multirow{2}{*}{1.19 }  & \multirow{2}{*}{10.93} & \multirow{2}{*}{1.11}   \\ 
$p_{Tj}>30$~GeV,  $|\eta_j|<2.5$, $n_j\ge 2$, $n_l= 2$ &&&\\
two extra b-tags (four in total) & 0.43 &  4.21 & 0.21\\ 
$|m_{H}^{\text{BDRS}}-m_H|<10$~GeV, $m_{b\bar{b}}>110$~GeV  & 0.077 &  0.111 & 0.003 \\ 
\hline
$m_{\ell\ell}>75$~GeV  & 0.056 &  0.082 & 0.003 \\ 
\end{tabular} 
\caption{Cut-flow for signal and backgrounds at LHC ${\sqrt{s}=13~\text{TeV}}$. The selection follows the BDRS
analysis described in the text. Rates are in $fb$ and account for  70\%(1\%) $b$-tag(mistag) rate, hadronization, 
and underlying event effects.}
\label{tab:cut_flow}
\end{table}
%-------------------------------------------------------

As for the $t\bar{t}H$ analysis with hadronic top decays $S/B < 1$~\cite{tth_plehn}. The bounds can be improved by accounting for the signal and background distribution profiles. We  use the two dimensional distribution $(p_{TH}, \Delta \phi_{jj})$ for our log-likelihood test. The $p_{TH}$ distribution
drops  slower for signal than for the continuum background. This is the main reason to look at the boosted kinematics for this signal.  In addition, the
azimuthal  angle between the two leading jets $\Delta \phi_{jj}$ (either $b$-tagged or not) presents a different profile thanks to the different radiation
profiles of signal and background. In Fig.~\ref{fig:luminosity} we present the expected 95\% CL limit on the signal strength $\sigma/\sigma^{SM}$ in the 
dileptonic $t\bar{t}H$ channel as function of the LHC luminosity. Sensitivity to the SM coupling will require $\sim 175$~fb$^{-1}$ of integrated luminosity Additional  improvements for the signal extraction can be achieved, {\it e.g.}, via the matrix element method or a neural network~\cite{tth_cms,tth_atlas}.

%----------------------------------------------------------------------------------------
\begin{figure}[b!]
\centering
\includegraphics[width=0.85\columnwidth]{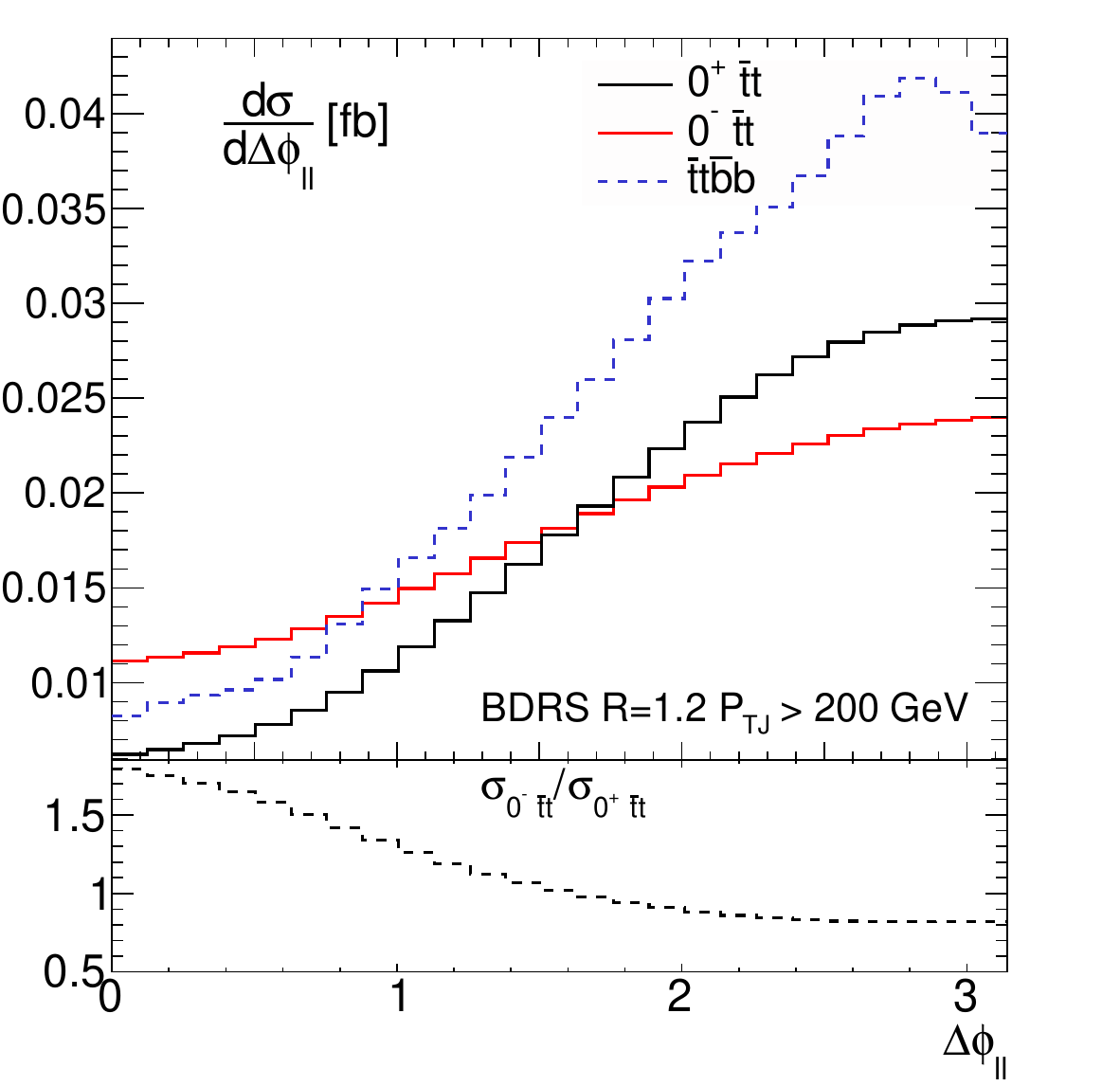}
\caption{Azimuthal correlation between the two leptons  $\Delta \phi_{\ell\ell}$ calculated
in the lab-frame after the BDRS analysis.}
 \label{fig:dphill_bdrs}
\end{figure}
%----------------------------------------------------------------------------------------

%----------------------------------------------------------------------------------------
%\begin{figure}[b!]
%\centering
%\includegraphics[width=0.87\columnwidth]{./CLs_dphill.pdf}
%\caption{Expected statistical significance to distinguish CP-even and CP-odd $t\bar{t}H$ couplings 
%in the dileptonic channel (cut flow from Table~\ref{tab:cut_flow}). The $\Delta\phi_{\ell\ell}$ 
%variable introduced in this Letter (dashed) is compared with $\Delta\phi_{\ell\ell H}$ of Ref.~\cite{Boudjema:2015nda} (dotted).} 
% \label{fig:luminosityCP}
%\end{figure}
%----------------------------------------------------------------------------------------

%----------------------------------------------------------------------------------------
\begin{figure}[t!]
\centering
\includegraphics[width=0.95\columnwidth]{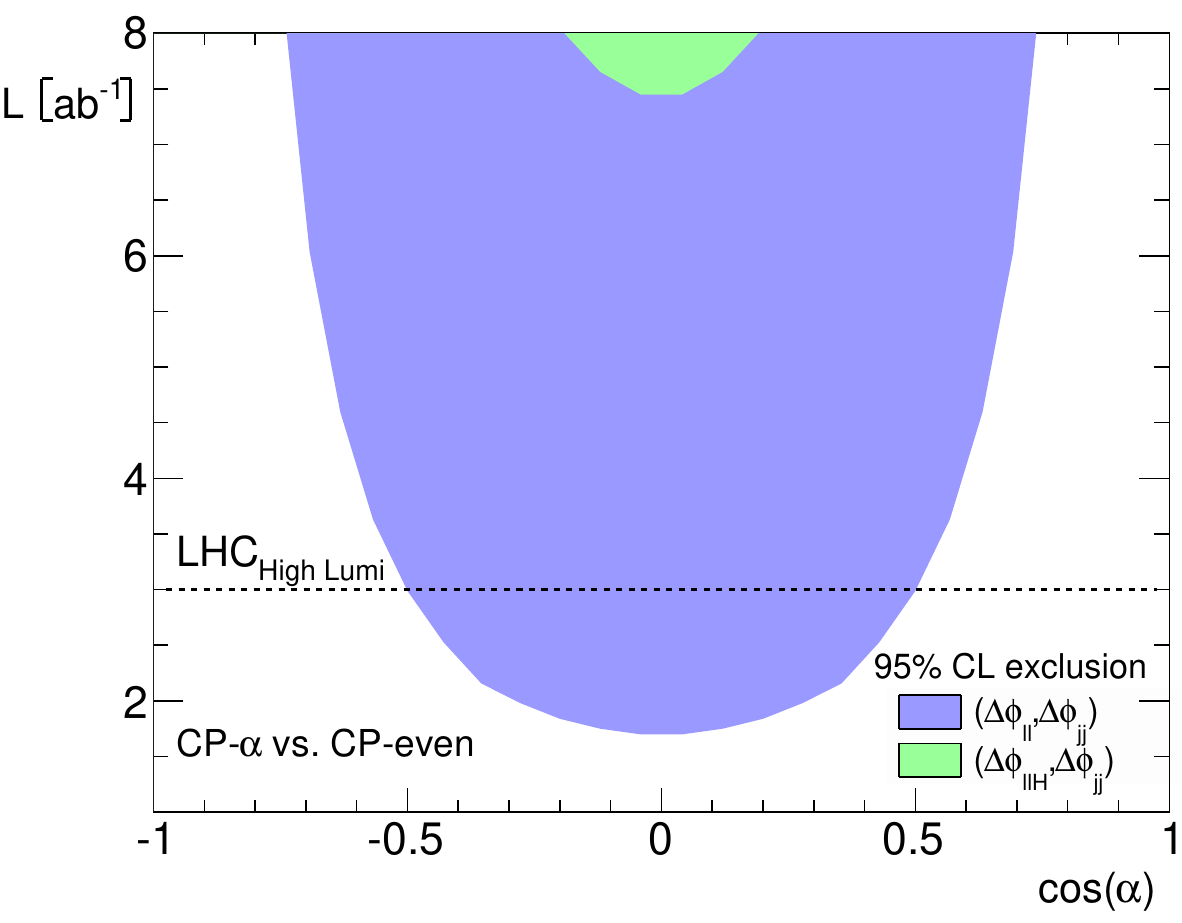}
\caption{Luminosity required to distinguish CP-even $t\bar{t}H$ coupling  from couplings with arbitrary CP-phase.}
 \label{fig:alpha_lumi}
\end{figure}
%----------------------------------------------------------------------------------------

Next we consider CP discrimination in the Higgs-top coupling. We further
require the dilepton invariant mass to be $m_{\ell\ell}>75$~GeV; enhancing the
sensitivity of $\Delta \phi_{\ell \ell }$ from $\sigma_{0^-t\bar{t}}/\sigma_{0^+t\bar{t}}\sim1.4$ to 
$\sim1.9$ at $\Delta \phi_{\ell \ell }\sim0$. After all cuts, the CP-even and CP-odd distributions 
of $\Delta\phi_{\ell\ell}$ (and $t\bar{t}b\bar{b}$ background) are shown in  Fig.~\ref{fig:dphill_bdrs}. 
Note that this remains sensitive to the Higgs-top CP-structure after a realistic  simulation that includes 
in particular NLO QCD effects.

To analyze $\Delta\phi_{\ell\ell}$'s discriminating power, we perform a binned log-likelihood test in 
$(\Delta \phi_{\ell \ell},\Delta \phi_{jj})$. To focus only on measurement of $\alpha$, we fix the number of signal events
 to the SM prediction.  In Fig.~\ref{fig:alpha_lumi}, we plot the expected statistical  significance
with which this analysis can distinguish a top-Higgs coupling with arbitrary CP-phase from the CP-even $\alpha = 0$ case.  
As can be seen, 95\% CL exclusion of the CP-odd case should be possible with $\sim1.8~$ab$^{-1}$ of data, and the high luminosity 
LHC would be able to distinguish the CP-even couplings from couplings with $|\cos\alpha|\lesssim 0.5$. 
This bound can be further improved by using more observables in our likelihood test and by including the three $b$-tag  sample.
%that, despite of presence of the $t\bar{t}jj$ background, should still be able to add significant statistical power~\cite{tth_cms,tth_atlas}. 

In Fig.~\ref{fig:alpha_lumi}, we also compare our analysis with another lab-frame observable proposed in Ref.~\cite{Boudjema:2015nda}.
Here the  angle is defined around the Higgs axis: $\Delta\phi_{\ell\ell H}$. 
We notice that the CP sensitivity of this observable decreases in the boosted regime in comparison with $\Delta\phi_{\ell\ell}$.\medskip

%----------------------------------------------------------------------------------------
%\section{Summary}

In this Letter, we have introduced a simple lab-frame variable, $\Delta\phi_{\ell\ell}$, which can be used to measure the CP-properties 
of the top-Higgs coupling in the dileptonic channel. On theoretical grounds, we expect this variable to be most useful when the Higgs
is significantly boosted, which pushes us into a kinematic regime where significant reductions in background can be obtained via 
substructure tagging when $H$ decays to $b\bar{b}$. The high-$p_{TH}$ kinematic regime, where  $\Delta\phi_{\ell\ell}$  is most sensitive 
to CP, also lends itself to a boosted Higgs analysis, which can be used to significantly enhance the discovery potential of the $t\bar{t}H$ channel.
We show a detailed theoretical study at NLO in the four $b$-tag sample, demonstrating that the LHC with 
$\sqrt{s}= 13$~TeV should be capable of probing the SM-strength top-Higgs coupling with $\sim175$~fb$^{-1}$, and then 
distinguishing between the CP-even and CP-odd couplings with  $\sim1.8$~ab$^{-1}$. Improvements may be possible, for example by
including the three $b$-tag sample, or adding additional discriminating variables. \medskip

%----------------------------------------------------------------------------------------
\begin{acknowledgments}
\textbf{Acknowledgments} -- We would like to thank Frank Krauss, Silvan Kuttimalai and Philipp Maierhoefer for valuable conversations.
\end{acknowledgments}

%----------------------------------------------------------------------------------------

\end{document}